\documentclass{article}

\usepackage{arxiv}   % template for arxiv
\usepackage{float}
\usepackage{amsmath}
\usepackage{booktabs}
\usepackage{multirow}
\usepackage{graphicx}
\usepackage{threeparttable}
\usepackage{wasysym}   % symbols and icons
\usepackage[normalem]{ulem}

\usepackage{url}

\usepackage{subcaption}

\usepackage{makecell}  % multi line cell table
\usepackage[T1]{fontenc}   % make sure UTF-8 box-drawing chars print cleanly
\usepackage{dirtree}  % directory tree

\title{OpenCAMS: An Open-Source Connected and Automated Mobility Co-Simulation Platform for Advancing Next-Generation Intelligent Transportation Systems Research}

\author{
    Minhaj Uddin Ahmad \\
    The University of Alabama\\
    Tuscaloosa, AL\\
    \texttt{mahmad12@crimson.ua.edu} \\
   \And
    Akid Abrar \\
    The University of Alabama\\
    Tuscaloosa, AL\\
    \texttt{aabrar@crimson.ua.edu} \\
   \And
    Sagar Dasgupta \\
    The University of Alabama\\
    Tuscaloosa, AL\\
    \texttt{sdasgupta@ua.edu} \\
   \And
    Mizanur Rahman \\
    The University of Alabama\\
    Tuscaloosa, AL\\
  \texttt{mizan.rahman@ua.edu} \\
}

\begin{document}
\maketitle
%As connected and automated mobility systems become increasingly complex, simulation platforms require to evolve for support integrated modeling across roadway traffic behavior, perception systems, and vehicle-to-everything (V2X) communication. Traditional simulators are typically designed with a narrow focus—either traffic flow, sensor-level autonomy, or network dynamics—making them unsuitable for evaluating connected transportation system applications holistically. To address this gap, By enabling plug-and-play experimentation across layers of traffic, perception, and communication, CAM-OpenSim empowers researchers to investigate performance, safety, security, and sustainability with unprecedented interoperability and depth. 

\begin{abstract}

We introduce OpenCAMS (\textbf{Open}-Source \textbf{C}onnected and \textbf{A}utomated \textbf{M}obility Co-\textbf{S}imulation Platform), an open-source, synchronized, and extensible co-simulation framework that tightly couples three best-in-class simulation tools: (i) Simulation of Urban Mobility (SUMO), (ii) Car Learning to Act (CARLA), and (iii) Objective Modular Network Testbed in C++ (OMNeT++). OpenCAMS is designed to support advanced research in transportation safety, mobility, and cybersecurity by combining the strengths of each simulation domain. Specifically, SUMO provides large-scale, microscopic traffic modeling; CARLA offers high-fidelity 3D perception, vehicle dynamics, and control simulation; and OMNeT++ enables modular, event-driven network communication, such as cellular vehicle-to-everything (C-V2X). OpenCAMS employs a time-synchronized, bidirectional coupling architecture that ensures coherent simulation progression across traffic, perception, and communication domains while preserving modularity and reproducibility. For example, CARLA can simulate and render a subset of vehicles that require detailed sensor emulation and control logic; SUMO orchestrates network-wide traffic flow, vehicle routing, and traffic signal management; and OMNeT++ dynamically maps communication nodes to both mobile entities (e.g., vehicles) and static entities (e.g., roadside units) to enable C-V2X communication. While these three simulators form the foundational core of OpenCAMS, the platform is designed to be expandable and future-proof, allowing additional simulators to be integrated on top of this core without requiring fundamental changes to the system architecture. The OpenCAMS platform is fully open-source and publicly available through its GitHub repository \url{https://github.com/minhaj6/carla-sumo-omnetpp-cosim}, providing the research community with an accessible, flexible, and collaborative environment for advancing next-generation intelligent transportation systems.

\end{abstract}

\section{Background and Motivation}
Modern transportation systems are undergoing a paradigm shift driven by the convergence of computing, communication, and automation technologies. At the forefront of this transformation are Intelligent Transportation Systems (ITS), which aim to enhance road safety, reduce congestion, and improve overall mobility by leveraging real-time data, decentralized control, and distributed intelligence. ITS architectures are increasingly evolving into Transportation Cyber-Physical Systems (TCPS), where computational elements are deeply embedded within physical transportation infrastructure and vehicles. These systems enable feedback-driven control and decision-making to optimize traffic flow, reduce crashes, minimize fuel consumption, and lower emissions.

Complementing the TCPS paradigm is the emergence of Digital Twin (DT) technologies, which offer real-time virtual replicas of physical systems that evolve in parallel with their real-world counterparts. DTs support predictive analytics, dynamic modeling, and scenario testing for improved planning, operations, and system resilience. While Internet of Things (IoT) technologies provide the sensing and connectivity backbone, it is the integration of IoT with TCPS and DT that enables actionable intelligence in complex transportation environments. Collectively, these technologies form a multi-layered digital ecosystem designed not only to optimize mobility and safety but also to ensure continuity of operations in the face of uncertainties.

Together, ITS, TCPS, DT, and IoT hold immense promise for improving transportation outcomes. However, the growing complexity and interconnectedness of these systems also introduce substantial cybersecurity risks. As vehicles and infrastructure increasingly depend on software-defined control and wireless communication, they become susceptible to a wide range of attack vectors, ranging from signal spoofing and jamming to message injection, denial-of-service (DoS) attacks, and coordination disruption. These vulnerabilities not only jeopardize safety and mobility but can also compromise public trust in intelligent transportation technologies. As such, cybersecurity has emerged as a fundamental pillar of next-generation ITS research, on par with safety and efficiency.

Central to the operation and effectiveness of modern ITS and TCPS architectures is connectivity. Technologies such as Cellular Vehicle-to-Everything (C-V2X) enable real-time information sharing between vehicles, infrastructure, and cloud services. C-V2X supports both PC5 sidelink for direct, low-latency communication and Uu interface over 4G/5G networks for broader coverage and cloud integration. These capabilities are indispensable for latency-sensitive and cooperative applications such as platoon control, intersection coordination, cooperative perception, and dynamic traffic signal control. Today, connectivity is no longer optional, it is foundational. It enables not just functional capabilities but also determines the system's adaptability and resilience. However, this reliance on connectivity further enlarges the attack surface for adversaries, who may exploit vulnerabilities at the physical, communication, or application layers. Consequently, testing the cyber resilience of connected transportation systems has become an urgent research priority. Evaluating these systems under real-world cyber-physical attack scenarios is critical but fraught with logistical, financial, and regulatory challenges.

Real-world experimentation involving connected vehicles, autonomous agents, and cyberattack scenarios is expensive, difficult to scale, and potentially hazardous. Deploying full-scale testbeds with adversarial interactions—such as spoofing, jamming, or malicious behavior injection—can compromise safety or violate regulatory constraints. Moreover, reproducing edge cases or rare-event scenarios in real-world settings is often infeasible. Simulation thus emerges as a critical tool in the development pipeline for intelligent and secure transportation systems. It enables researchers to conduct functional testing, stress testing, and threat modeling in a controlled, repeatable, and safe environment.

Despite the importance of connectivity, there remains a significant gap in the ability to simulate these systems in a holistic and integrated manner. Existing simulation tools tend to focus on specific subsystems in isolation. For instance, traffic simulators such as Simulation of Urban MObility (SUMO)~\cite{SUMO2018} or Vissim~\cite{ptv_vissim} are well-suited for modeling large-scale traffic dynamics but lack support for high-fidelity perception or real-time communication. Driving simulators like Car Learning to Act (CARLA)~\cite{Dosovitskiy17} lack standard microscopic traffic simulation models and are often decoupled from realistic traffic and networking environments. Network simulators like Objective Modular Network Testbed in C++ (OMNeT++)~\cite{omnetpp} or Network Simulator 3 (ns-3)~\cite{ns3website} are capable of modeling communication protocols but assume either static or abstract mobility models and cannot accurately capture the real-time dynamics of vehicle behavior.

These siloed approaches limit the ability to evaluate closed-loop interactions between vehicle mobility, sensing, decision-making, and connectivity. This limitation becomes particularly problematic when assessing scenarios that involve mixed autonomy, cooperative behavior, or cybersecurity threats, where the interplay between perception, planning, and communication is critical. Furthermore, many existing frameworks are not designed to be modular, extensible, or open-source, making it difficult for researchers to experiment with new algorithms, protocols, or coordination strategies.There is a clear need for a simulation framework that can support such evaluations by tightly integrating traffic, autonomy, and communication components, alongside cybersecurity threat injection and mitigation modeling.

To address this gap, we present OpenCAMS, a co-simulation framework that unifies three domain-specific simulators into a tightly synchronized simulation environment: SUMO~\cite{SUMO2018} for microscopic traffic simulation, CARLA~\cite{Dosovitskiy17} simulator for physics-based 3D environment and sensor simulation, and OMNeT++ simulator for full protocol stack simulation of C-V2X from application layer to MAC layer. Each simulator was selected based on its technical capabilities, extensibility, and community support. SUMO is open-source, lightweight, and scalable, capable of simulating thousands of vehicles and driving behavior. While commercial simulators like PTV Vissim~\cite{ptv_vissim} or Aimsun~\cite{aimsun2025} offer advanced behavior models, their closed-source nature and licensing restrictions limit extensibility and co-simulation integration. 

CARLA~\cite{Dosovitskiy17}, based on Unreal Engine, supports full-stack autonomous driving simulation with configurable weather, lighting, and sensor models (e.g., LiDAR, camera, radar). Alternatives, such as LGSVL~\cite{rong2020lgsvl} or AirSim~\cite{rong2020lgsvl} are deprecated and have been officially discontinued. OMNeT++, paired with frameworks like INET and Veins, supports the modeling of DSRC, C-V2X, LTE, and 5G communication stacks. Unlike ns-3, which focuses more on low-level protocols, provides limited modularity and visualization, OMNeT++ offers fully modular implementations of network stacks and applications with INET Veins, etc. Through Veins framework, it seamlessly integrates with SUMO via Traffic Control Interface (TraCI), making it more suitable for closed-loop vehicular communication testing. OMNeT++ also provides it own Network Description (NED) language for extensible network descriptions, eclipse based integrated development environment.

By autonomy, we refer to the SAE levels of driving automation from Level 0 to Level 5, which range from fully human-driven vehicles with no driving automation (Level 0) to fully autonomous vehicles or full driving automation (Level 5)~\cite{sae2014taxonomy}. Hence, OpenCAMS supports the simulation of connected vehicle environments where all vehicles are connected, regardless of their autonomy level. This includes scenarios where all vehicles are connected and human-driven, a mix of connected human-driven and autonomous vehicles (AV), or fully connected autonomous fleets, all within a unified, extensible co-simulation platform.

OpenCAMS executes these simulators in a synchronized loop using discrete time-step simulation. Traffic flow updates from SUMO, ego vehicle control and sensing from CARLA, and C-V2X message exchanges in OMNeT++ all occur within each time tick, ensuring consistent state evolution and minimizing causality violations. The system can be extended to include additional clients, such as Global Navigation Satellite System (GNSS) simulators for spoofing analysis, Autoware~\cite{autoware2025} stacks for autonomous control logic, or Python scripts to manipulate infrastructure like traffic lights or pedestrian behavior. The key featues of OpenCAMS are:

\textbf{Integration-Driven:} OpenCAMS establishes bidirectional, time-synchronized coupling across all three simulators, capturing realistic interplay between mobility, autonomy, and communication in a closed-loop environment.

\textbf{Open-Source Component-Based:} All simulators used in OpenCAMS—SUMO, CARLA, and OMNeT++, are freely available and widely supported by active research communities, ensuring accessibility and long-term maintainability.

\textbf{Generalized and Expandable:} While the core simulation loop depends on SUMO, CARLA, and OMNeT++, OpenCAMS supports attaching additional simulation tools, including GNSS signal simulators, Autoware or Robot Operating System (ROS) based stacks, and lightweight Python clients for dynamic control (e.g., traffic light manipulation via SUMO’s TraCI interface).

The remainder of this paper is structured as follows. Section~\ref{sec:relevant-works} presents a review of existing co-simulation platforms, highlighting their limitations and motivating the need for OpenCAMS. Section \ref{sec:method} details the architecture of OpenCAMS, including its discrete time-step synchronization strategy and coupling mechanisms between SUMO, CARLA, and OMNeT++. Subsections ~\ref{sim-setup}, ~\ref{running-cosim}, and ~\ref{data-collection} outlines the simulation setup process, including map preparation, synchronization routines, and system-level integration procedures. Section \ref{sec:applications} discusses a comprehensive set of connected and automated mobility research enabled by OpenCAMS, each requiring the interplay of microscopic traffic modeling, high-fidelity perception, and C-V2X communication. Finally, Section \ref{sec:conclusions} concludes the paper and outlines directions for future research and extension of the framework.

\section{Significance of OpenCAMS}
\label{sec:relevant-works}

The development and evaluation of intelligent and connected transportation system technologies depend heavily on simulation tools that can capture the complexity of real-world traffic, environmental perception, and communication dynamics. Historically, these simulators have been built to specialize in individual domains such as traffic modeling, perception, or networking. While effective in isolation, the lack of integration across these domains has limited their utility for conducting comprehensive connected and automated mobility research. This section surveys the current state of individual and co-simulation tools, examines their strengths and limitations, and highlights the research gap addressed by our integrated platform.

One of the most widely used roadway traffic simulators in research and academia is the SUMO~\cite{SUMO2018}. It is a highly scalable, open-source microscopic traffic simulator capable of handling thousands of vehicles in large transportation roadway networks. SUMO supports various microscopic traffic flow models, such as Intelligent Driver Model (IDM)~\cite{treiber2000congested} and Krauss~\cite{kraussdeutsches} car-following models, as well as several lane changing models, such as DK2008~\cite{krajzewicz2010trafficDK2008}, LC2013~\cite{erdmann2015sumoLC2013}, and SL2015~\cite{sumoSL2015}.  LC2013, developed by Jakob Erdmann, builds upon the earlier DK2008 model by Daniel Krajzewicz and is widely used for roadway traffic simulation. For simulations that utilize sub-lane resolution for allowing lateral movement inside the lane, the SL2015 model is activated. SUMO enables simulation of various aspects of roadway traffics, such as route assignment, vehicle interactions, traffic signal control, and intermodal traffic. SUMO's key strengths are its speed, openness, and powerful APIs, such as Traffic Control Interface (TraCI)~\cite{wegener2008traci}, which make it ideal for integration with other types of simulators, such as communication and robotic simulators. However, SUMO's simplified vehicle dynamics and lack of 3D environmental representation limit its use for testing perception-based algorithms or human-in-the-loop simulations. Commercial microscopic traffic simulators, such as PTV Vissim~\cite{ptv_vissim} and Aimsun~\cite{aimsun2025} offer advanced capabilities for multimodal traffic simulation, pedestrian modeling, and traffic signal optimization. While Vissim excels in signal timing and behavioral modeling, Aimsun offers both microscopic and mesoscopic simulation scales. These features are not exclusive to commercial traffic simulators; and are available in open-source alternative SUMO. While SUMO may lack the user-friendly graphical interfaces offered by commercial tools for certain tasks, it remains a powerful and flexible option for research. The use of commercial microscopic simulators is often limited by high licensing costs, lack of open-source extensibility, and limited integration with external simulators.

As perception and autonomy become critical components of ITS, simulators, such as CARLA~\cite{Dosovitskiy17}, LG Silicon Valley Lab simulator (LGSVL)~\cite{rong2020lgsvl}, and AirSim~\cite{airsim2017fsr}, have gained popularity. They all feature a 3D model of an environment where the components utilize a sophisticated physics engine for realistic simulation. These simulators leverage existing game engines, such as Unity and/or Unreal Engine, which provide mechanisms for creating photorealistic 3D environments, simulating complex physics, and supporting sensor modeling. CARLA, built on Unreal Engine, is an open-source autonomous driving simulator designed to provide high-fidelity simulation of sensor data including LiDAR, cameras, radar, and GPS. Users can configure environmental conditions, such as lighting and weather, in CARLA. Additionally, CARLA supports synchronous time-stepped simulation, which is essential for precise integration with other simulators and/or tools.  LGSVL simulator, based on Unity and originally designed to support AV software stack of Apollo~\cite{apollo2025} and Autoware~\cite{autoware2025}. LGSVL provides a similar set of features as CARLA; however, its development was suspended in 2022, leaving CARLA as the de facto open-source standard for high-fidelity vehicle and sensor simulation. AirSim, developed by Microsoft Research and based on the Unity game engine, was originally designed for drone simulations but later expanded to include ground vehicles. However, since AirSim is primarily geared toward drones, the support for ground vehicle-related applications is limited. Microsoft stopped supporting AirSim development and maintenance in 2023. All of these AV simulators are widely used in reinforcement learning and AI research due to their capability for controlling vehicles and generating realistic synthetic data through the sensors. However, they lack the ability (i) to simulate large-scale roadway traffic using standardized microscopic traffic flow models, (ii) to integrate a communication stack, such as C-V2X, which limits their applicability in connected transportation systems. We choose CARLA for simulating the 3D environment and perception due to its ongoing active development and wide adoption in academic and industry-based research.  

Native support for SUMO–CARLA co-simulation was introduced in CARLA version 0.9.8, enabling bidirectional synchronization between CARLA and the SUMO microscopic traffic simulator using an extended TraCI interface. This integration allows for hybrid traffic and perception simulation, where CARLA can manage high-fidelity vehicles with sensors, and SUMO orchestrates surrounding roadway traffic and traffic signal control logic. Several studies have extended this baseline to build functional co-simulation frameworks. Li et al.~\cite{li2021novel} developed a traffic simulation framework in which CARLA is used for detailed vehicle simulation and SUMO for traffic-level interactions and control, enabling synchronized multi-vehicle AV testing. Cantas and Guvenc~\cite{cantas2023customized} presented a customizable co-simulation environment that leverages CARLA for autonomous vehicle control development and SUMO for traffic generation and enabling reinforcement learning-based decision-making. Roccotelli et al.~\cite{roccotelli2024co} proposed a co-simulation platform that integrates CARLA and SUMO for evaluating autonomous mobility in mixed traffic contexts, showcasing a modular architecture to support realistic AV testing in urban environments. While these platforms successfully bridge the gap between traffic modeling and perception, they generally lack support for C-V2X communications, which enables wireless connectivity. Moreover, vehicle-to-vehicle (V2V) communications are often simulated with simplistic assumptions and do not model the standard protocol stacks for communication networking.  

The emergence of connected transportation system technologies has driven the need for simulating C-V2X communication. OMNeT++~\cite{omnetpp} is a discrete-event, modular simulation platform widely used for modeling communication networks. With frameworks, such as INET~\cite{inet_framework} and Veins~\cite{veins2025}, OMNeT++ can simulate wireless protocols, including Dedicated Short-Range Communications (DSRC), Long Term Evolution (LTE), and 5G. Its strengths include its visual debugging interface, modular architecture, and extensibility. ns-3~\cite{ns3website}, while offering a more detailed and lower-level simulation of network protocols, does not feature the extensive modules that OMNeT++ does. ns-3 also lacks advanced visualizations and visual debugging compared to those offered by OMNeT++. On the other hand, OMNeT++ provides a rich graphical interface, an integrated development environment, and its own extensible NED (NEtwork Description) programming language along with C++. Both communication simulators are capable, but OMNeT++ is preferred in this co-simulation context due to its flexibility and compatibility with SUMO via TraCI, leveraging the integration between SUMO and OMNeT++ that is facilitated by Veins (i.e., Vehicles in Network Simulation)~\cite{veins2025}.

Veins provides a robust integration framework for coupling SUMO and OMNeT++, enabling bidirectional and time-stepped co-simulation. Through the TraCI interface, vehicle mobility and network simulation are synchronized, allowing researchers to evaluate C-V2X communication under realistic roadway traffic conditions. Veins also includes features for modeling signal attenuation using building polygons. However, Veins lacks any notion of vehicle-level sensor simulation since it utilizes modes (both fixed and mobile nodes) provided by SUMO. It is limited by the node types SUMO has. However, for advanced autonomous driving, vehicle control and perception, as well as Vehicle-to-Vehicle (V2V) and Vehicle-to-Infrastructure (V2I) connected vehicle applications, sensors provided by simulators like CARLA are necessary.

To overcome this, several co-simulation frameworks have been proposed. Veins-Carla~\cite{hardes2023poster} replaces SUMO with CARLA to incorporate high-fidelity vehicle dynamics into the Veins pipeline. While this integration improves realism by leveraging CARLA’s high-fidelity rendering and sensor models, scalability depends on how CARLA is configured. When all vehicles are rendered with full sensor stacks and visuals, performance may degrade; however, CARLA supports scalable configurations through headless mode, sensor data throttling, and selective rendering. Headless mode refers to running a software without a graphical user interface (GUI), which significantly reduces computational overhead by skipping visual rendering. Since CARLA replaced the SUMO for managing traffic, it loses the flexibility of using standard microscopic traffic flow models with configurable parameters. MS-VAN3T-CARLA \cite{carletti2024ms} bridges CARLA and ns-3 using a gRPC-based middleware, running all vehicles within CARLA. gRPC or Google Remote Procedure Call is google's implementation of Remote Procedure Call (RPC). RPC is a software protocol where one software, e.g., ns-3, can invoke a function in a different software e.g., CARLA. The two software involved here may reside in the same computer or a different computer in a different network. This gRPC-based middleware enables communication between ns-3 and CARLA to support photorealistic cooperative perception testing and realistic C-V2X communication, though the scalability and runtime performance are determined by simulation design choices and hardware provisioning rather than inherent limitations of the simulator itself. MS-VAN3T-CARLA also has a dependency on OpenCDA which brings design constraints as it is geared toward co-operative driving automation (CDA) and additional dependencies. On the other hand, CARLA-SUMO-Artery \cite{bouchouia2022simulator} integrates CARLA and SUMO with the Artery communication stack to support ETSI ITS-G5 protocols, and is primarily designed for simulating security scenarios related to cooperative and automated driving. While effective for prototyping misbehavior injection and V2X message tampering, the framework lacks support for alternative communication standards of C-V2X. Furthermore, although SUMO and OMNeT++ are synchronized via Artery’s fixed time-step mechanism, CARLA is integrated asynchronously through a Python bridge, resulting in partial synchronization that may cause temporal misalignment in closed-loop scenarios requiring tight coupling between perception and communication modules.

A particularly notable recent framework is OpenCDA (Open Cooperative Driving Automation)~\cite{xu2021opencda}, which provides a unified and open-source platform for testing cooperative driving strategies, such as platooning, merging, and coordinated lane changes. OpenCDA supports both centralized and decentralized coordination architectures, integrating CARLA for vehicle-level simulation and incorporating SUMO for scalable network-wide background traffic generation. The framework includes built-in modules for cooperative behavior logic, perception fusion, and inter-vehicle communication. Although the authors claim that OpenCDA is compatible with external network simulators, such as ns-3, no detailed methods or implementation for such integration are provided. As a result, its ability to simulate realistic communication-layer disruptions, protocol-level interactions, or advanced cybersecurity scenarios remains limited compared to frameworks with full C-V2X stack integration. The framework of OpenCDA requires the use of all algorithms needed for cooperative driving automation. However, a connected vehicle application may not require cooperative driving components. On the other hand, Simutack~\cite{simutack:2023} integrates all three simulators, i.e., SUMO, CARLA and OMNeT++, like this study. However, in their architecture, the data generated from the simulators directly feeds into a cybersecurity attack generation framework, which makes it difficult to use other use cases that are not covered by the attack generation framework. They provide a specific web interface to interact with the simulators. This abstraction layer limits the use cases outside what the authors intended. A notable modular alternative is Eclipse MOSAIC~\cite{schrab2022MOSAIC}, which functions as a co-simulation middleware connecting domain-specific simulators through dedicated adapters for each simulator. While offering flexible and standardized interface between individual simulators, the CARLA adapter is still under development. MOSAIC Extented, a commercial version of MOSAIC provides support of 3D environment and vehicle physics simulation through a proprietary simulator PHABMACS~\cite{phabmacs}. However, PHABMACS restricts environmental realism by procedurally generating minimal road furniture (no curbs, lamp posts, or benches), and targets small-scale ADAS prototyping—trading off high-detail physics for fewer vehicles support. CARLA provides more overarching scenarios with more vehicle options and environmental customization, such as weather condition. Moreover, PHABMACS lacks open interfaces comparable to CARLA’s extensive tooling and community support. Although MOSAIC Extended has a CARLA adapter currently in internal development, it may be exclusive to the commercial tier. Once released, MOSAIC Extended could provide functionality comparable to the OpenCAMS co-simulation framework presented in this study.

\begin{table}[H]
\centering
\footnotesize

\newcommand*\feature[1]{\ifcase#1 \Circle\or\LEFTcircle\or\CIRCLE\fi}
\newcommand*\f[4]{\feature#1 & \feature#2 & \feature#3 & \feature#4}

\begin{threeparttable}
\caption{Comparison of co-simulation platforms}
\label{tab:comparison-table}
\begin{tabular}{@{}l cccc@{}}
\toprule
Platform  & \makecell{Environment \& \\ Perception Sensor} & \makecell{Traffic \\ Simulation} & \makecell{Communication \\ Simulation} & Generalization \\
\midrule

Veins~\cite{sommer2011bidirectionally} & \f0220 \\
Veins-Carla~\cite{hardes2023poster} & \f2120 \\
ms-van3t-carla~\cite{carletti2024ms} & \f2121 \\
Carla-SUMO-Artery~\cite{bouchouia2022simulator} & \f2220 \\
PLEXE~\cite{segata2022multi} & \f0221 \\
OpenCDA~\cite{xu2021opencda} & \f2211 \\
Simutack~\cite{simutack:2023} & \f2220 \\
MOSAIC~\cite{schrab2022MOSAIC} & \f1222 \\
\textbf{Our platform} & \f2222\\

\bottomrule
\end{tabular}
\begin{tablenotes}[para, flushleft]
\item $\feature2=\text{full support}$
\item $\feature1 = \text{partial support}$ 
\item $\text{\feature0}=\text{no support}$
\end{tablenotes}
\end{threeparttable}
\end{table}

A consolidated comparison of major co-simulation platforms is presented in Table~\ref{tab:comparison-table}, evaluating each against three essential capabilities required for comprehensive Connected and Automated Mobility (CAM) simulation: high-fidelity environment and perception modeling, scalable microscopic traffic simulation, and flexible C-V2X communication support. The table uses intuitive markers to indicate whether each platform offers full, partial, or no support for these simulation capabilities. The generalization column indicates how comprehensive each co-simulation platform is in terms of CAM simulation capabilities. 

The comparison between these co-simulation frameworks highlights several research gaps. First, some of the co-simulation frameworks achieve a synchronous co-simulation, but are geared toward a specific application area. This makes it more challenging to repurpose the co-simulation platform for other CAM use cases. The abstraction provided by some existing co-simulation platforms is often a trade-off between ease of use and extent of generalization. Second, communication stacks are often tightly coupled to specific protocols, which reduces their generalizability. Third, synchronization mechanisms between simulators are either loosely implemented or rigid, limiting experimentation with diverse scenarios or additional simulation modules.

To address these critical challenges, we introduce a generalized and time-synchronized co-simulation platform that integrates three unique simulators for CAM applications: (i) SUMO for scalable traffic modeling, (ii) CARLA for high-fidelity environmental perception and vehicle dynamics, and (iii) OMNeT++ for flexible C-V2X communication simulation. Our platform employs a multi-client TraCI loop with strict time-step synchronization to ensure consistency across all three simulators. The unique feature of this co-simulation platform is that it avoids adding any abstraction layer on top of the simulators, preserving the flexibility and full potential of the underlying simulators. CARLA may simulate only a subset of vehicles requiring advanced in-vehicle sensors or roadside sensors, while SUMO handles the background roadway traffic. OMNeT++ dynamically maps communication nodes to vehicles and infrastructures, leveraging propagation models and customizable communication networking stacks. This design supports reproducible, modular, and extensible research across various domains, including the development and evaluation of reactive and proactive safety, mobility, and cybersecurity solutions relaated to CAM applications. In essence, our platform enables a unified, scalable, and realistic testbed for transportation research that bridges the gap between what each simulator is capable of in isolation and what modern connected and automated transportation systems demand.

%%%%%%%%%%%%%%%%%%%%%%%%%%%%%%%%%%%%%%%%%%%%%%%%%%%%%%%%%%%%

%%%%%%%%%%%%%%%%%%%%%%%%%%%%%%%%%%%%%%%%%%%%%%%%%%%%%%%%%%%%
\section{OpenCAMS Integration Approach}
\label{sec:method}

\begin{figure}[ht]  
    \centering
    \includegraphics[width=1\linewidth]{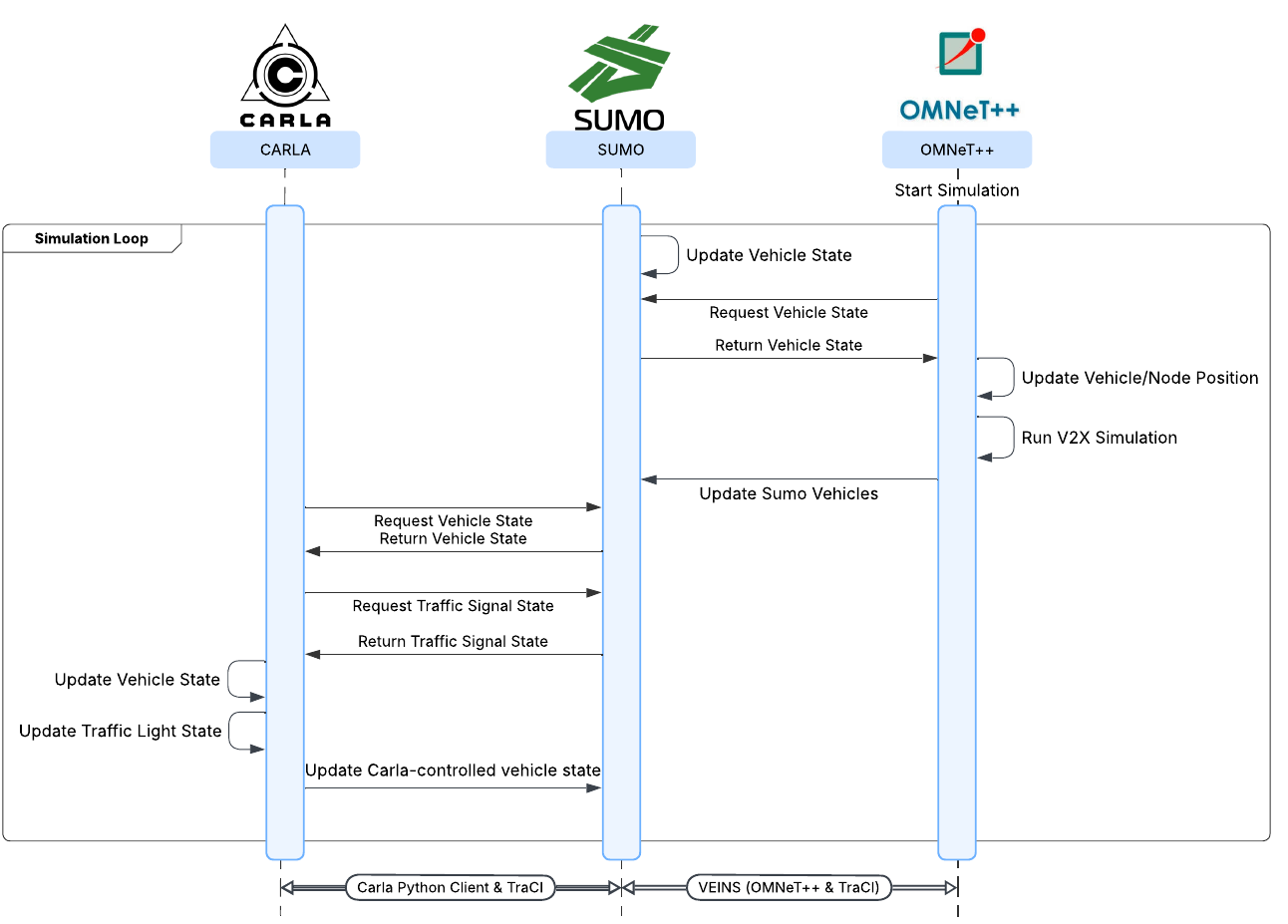}
    \caption{Simulation synchronization loop}
    \label{fig:sync-loop}
\end{figure}

For seamlessly integrating multiple simulation platforms in real-time, each simulator must be time-synchronous through synchronous discrete time-step simulation. Synchronous discrete-time-step simulation allows for granular control of each simulation step in each simulator and shares necessary information with the others. Each of the simulators (SUMO, OMNeT++, and CARLA) has the capability to run discrete-time step simulations. CARLA has support for both discrete synchronous time-step mode and asynchronous mode simulation, while SUMO and OMNeT++ only run in discrete time-step mode. This allows for creating bidirectional coupling mechanisms to share data between them at each simulation time step. The co-simulation between the simulators is achieved by two bidirectional coupling mechanisms as shown in Figure~\ref{fig:sync-loop}. CARLA and SUMO are coupled to share data in both directions. At each simulation step, vehicle positions and traffic light state are shared between CARLA and SUMO. This connection is bidirectional because the traffic signal controller and the vehicles can be controlled by either SUMO or CARLA. All the vehicles in the simulation can be mixed, where some are controlled by SUMO and others are controlled by CARLA. This bi-directionality provides the user with design flexibility to utilize the co-simulation platform for various research needs. For example, an AV with a full suite of sensors can be simulated using CARLA, and background traffic can be generated and controlled by SUMO, following specified traffic demand (i.e., Origin-Destination (OD) Matrix). The second coupling is with SUMO and OMNeT++, where the data are bidirectionally shared between them again. For SUMO and OMNeT++ coupling, we followed a similar approach as presented in Veins~\cite{sommer2011bidirectionally} with some modifications. Veins initializes a SUMO server that it uses for its Vehicular Ad Hoc Network (VANET) simulations. We modified Veins such that OMNeT++ can trigger the start of SUMO simulation with an adjustable $n$ number of SUMO clients. This creates the provisions to expose that instance of SUMO for further extensions, such as connecting to Carla, adding more TraCI clients to log data from SUMO or manipulating SUMO traffic in run-time. When multiple clients connect to SUMO, SUMO broadcasts the response to client request to all clients. We have added a logic such that OMNeT++ can ignore the client responses that is not meant for OMNeT++. The co-simulation progresses to the next simulation step when all the individual simulators are done with their respective tasks. Simulation of a C-V2X network protocols in OMNeT++ often takes more computational time than SUMO or CARLA. However, when OMNeT++ takes more computational time, SUMO and CARLA are frozen in simulation time until OMNeT++ completes its task. This holds for all involved simulators.For example, when CARLA takes more time to simulate data from computationally intensive sensors (e.g., LiDAR), the other simulators (SUMO and OMNeT++) are held frozen in simulation time. In this way, the synchrony and reproducibility of a co-simulation are ensured. However, the micro-freezes in simulation time to keep them synchronized at all times do not hamper the regular usage of the simulators. Figure~\ref{fig:sync-loop} is a sequence diagram that presents the flow in each simulation step. The series of events in the co-simulation environment is depicted in this sequence diagram, flowing from top to bottom. The steps for running the simulation, once all the necessary setup is complete, are described in Section~\ref {running-cosim}. When a simulation starts, SUMO populates the map with simulated vehicles that follow microscopic traffic simulation models. In SUMO, map information, the type of simulated vehicles (e.g., passenger cars and trucks), the car-following model (also known as the longitudinal driver behavior model), traffic control, and lane-changing model can be configured using XML file definitions. The section~\ref{sim-setup} provides an example for setting up the simulation and the prerequisites before running the simulation.

The co-simulation starts from the OMNeT++ Integrated Development Environment (IDE). When a simulation is launched, it triggers the start of an instance of the SUMO simulator. SUMO expects to receive two client-side connections using TraCI~\cite{wegener2008traci}. The first connection is from OMNeT++ itself, which uses our modified TraCIManager from Veins to connect as a client. When SUMO is started in multi-client mode, it waits for all clients to connect before it starts the simulation. Each client must specify their execution order to SUMO when connecting as a client. The modified TraCIManager in OMNeT++ is programmed in such a way that it sets itself up according to OMNeT++, which utilizes order 1. The second client is a Python script \verb|run_synchronization.py|, which tells SUMO to set its execution order as 2. This script is also responsible for connecting to CARLA and synchronizing vehicles and the traffic signal controller with CARLA. At first, SUMO runs the traffic simulation for one time step. Then the first client (OMNeT++) requests SUMO to report all the vehicle positions. OMNeT++ updates the location of its communication nodes corresponding to the vehicles according to the data received from SUMO. When OMNeT++ is done setting up the simulation and running the communication simulation for the time-step, it asks SUMO to proceed to the next simulation step. SUMO only proceeds to the next simulation step when all the connected clients ask SUMO to proceed to the next simulation step. SUMO waits for other client on CARLA side to synchronize - (i) request and receive the updated vehicle positions from SUMO to move all the vehicles, and (ii) change the traffic signal controller inside CARLA, simulate the CARLA-specific tasks, such as sensor simulation. Then, the CARLA side client also requests SUMO to proceed to the next simulation step. At this stage, since both the SUMO clients requested moving to the next simulation step, SUMO proceeds with the microscopic traffic simulation to the next step, updates the state information of vehicles and traffic signal controller . This completes the complete simulation loop, and it starts all over again for the next time step.

\subsection{Prerequisites of Co-Simulation Setup }
\label{sim-setup}

% add a flow for running the simulator
The CARLA simulator comes with several maps known as "Town Maps." Each map features different environments, such as, small suburban towns with bridges and rivers, urban town with residential and commercial buildings, map with roundabouts and large junctions, many-lane highways. These CARLA maps are defined in OpenDrive~\cite{dupuis2010opendrive} format, which is a popular HD map description format for simulation and Advanced Driver Assistance Systems (ADAS) systems. For demonstrating an example of co-simulation here, we use one of the existing CARLA towns. These existing CARLA towns provide a variety of environments that can cover a vast use cases, including urban, suburban, and rural scenarios. They can also be modified to add or remove custom map elements using Unreal Engine, which is a game engine CARLA is based on. Such maps can also be created from scratch using tools like RoadRunner~\cite{MathWorksRoadRunner}. The CARLA version 0.9.15 also introduces a procedural map generation pipeline to create a digital twin of a real-world location by pulling data from OpenStreetMap. All of these methods can result in a workable OpenDrive map definition ready for CARLA simulation. 

After a simulation map is ready for CARLA, the same map requires to be converted into a different format that SUMO can use. SUMO defines its simulation map as an XML file (e.g., roadnetwork.net.xml) that logically defines the roadway sections and landmarks. SUMO refers to their map as a 'net,' short for network file. SUMO provides a tool named \verb|netconvert.py| that can convert maps back and forth between formats. To ensure the best compatibility with co-simulation, CARLA also provides a utility program \verb|netconvert_carla.py|, which we recommend to use. This utility program can convert an OpenDrive (map.xodr) map definition to a SUMO network definition (map.net.xml). 

Once the map is ready, vehicle routes need to be defined. SUMO offers various utility programs to accomplish this. For example, the "duarouter" can take road traffic demand information as input and generate vehicle routes according to the provided demand definitions.

OMNeT++ can utilize the same map prepared for SUMO to provide a GUI representation through Veins~\cite{sommer2011bidirectionally}. However, to model the attenuation of the wireless C-V2X signal strength, OMNeT++ needs to know the locations of the obstacles in the map. When generating a map from the real world with the techniques above, these obstacle information can also be fetched from OpenStreetMap (OSM). A SUMO utility "polyconvert" has the ability to fetch this information from various sources (e.g., OSM and shape files) and export it as an XML file (map.poly.xml). Since CARLA maps may not necessarily represent a real-world location, we created a utility program that can connect to CARLA and export the locations of buildings as a SUMO polygon map (map.poly.xml). The utility program operates by requesting CARLA to provide bounding boxes for all city-level objects. Then the program filters the bounding box objects of type "building." Then, the utility program formats the data in XML format and saves it to local computer disk. However, the best results can be achieved by collecting the polygon data from OpenStreetMap if the location information is collected from the real world. The figure~\ref{fig:sumo-map} indicates the polygons using red color. When wireless C-V2X signals propagate through these objects, the signal will attenuate according to the signal attenuation model in OMNeT++. Note that an attenuation model adds the ability to estimate communication-specific delays and losses more realistically.

\subsection{Execution of co-simulation}
\label{running-cosim}
The steps to run the co-simulation platform with an example scenario are available at the project's GitHub repository. We will updated the documentation at the repository for any future changes and further developments. However, the steps for running the simulators will remain the same.

\begin{figure}[htb]
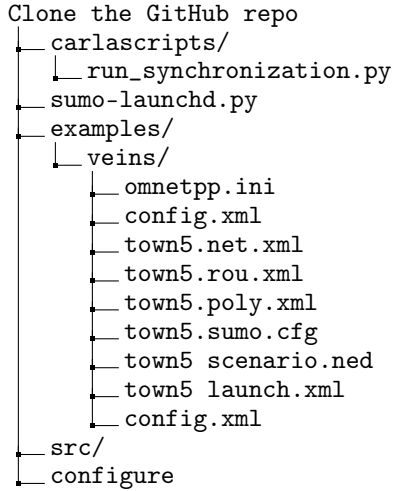

    \dirtree{%
    .1 Clone the GitHub repo.
    .2 carlascripts/.
    .3 run\_synchronization.py.
    .2 sumo-launchd.py.
    .2 examples/.
    .3 veins/.
    .4 omnetpp.ini.
    .4 config.xml.
    .4 town5.net.xml.
    .4 town5.rou.xml.
    .4 town5.poly.xml.
    .4 town5.sumo.cfg.
    .4 town5 scenario.ned.
    .4 town5 launch.xml.
    .4 config.xml.
    .2 src/.
    .2 configure.
    }
    \caption{Project file organization}
    \label{fig:file-tree}
\end{figure}

The necessary files that a user will change to adapt to a custom simulation scenario are shown in Figure~\ref{fig:file-tree}.

\begin{figure}[htbp]
    \centering
    \includegraphics[width=1\linewidth]{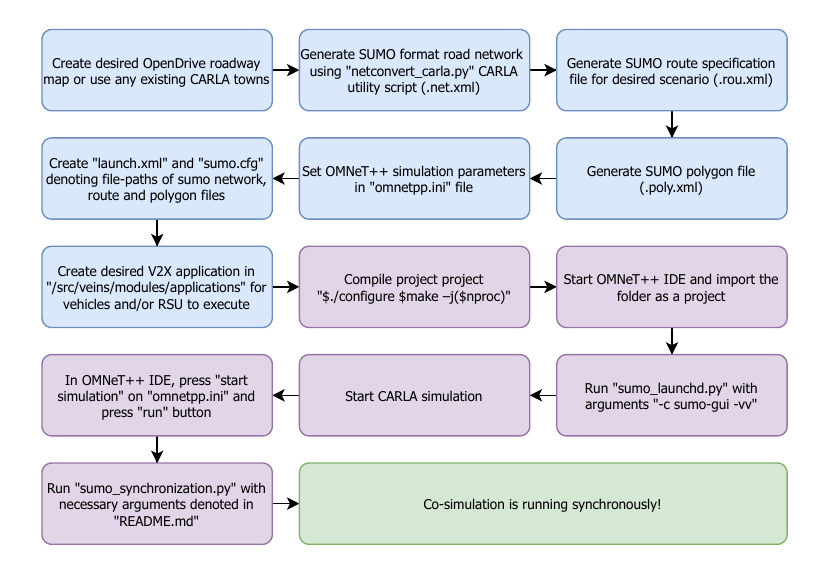}
    \caption{Workflow for customized simulation setup}
    \label{fig:flow-of-sim}
\end{figure}

Our GitHub repository contains a structured set of scripts and configuration files enabling co-simulation between CARLA, SUMO, and OMNeT++. The \verb|carlascripts/| directory contains Python scripts, such as \verb|run\_synchronization.py|, which manages time synchronization across simulators. This script accepts configurable parameters, and no changes is required in the provided scripts. The \verb|sumo-launchd.py| script serves as a daemon to launch and control SUMO, via TraCI. This script takes parameters, such as \verb|-n|, which specifies the number of clients SUMO should expect. No code changes are required to run the simulation. Within the \verb|examples/veins/| folder, a complete scenario is defined for the CARLA Town5 map, including SUMO network and route files (\verb|town5.net.xml|, \verb|town5.rou.xml|), visualization and configuration elements (\verb|town5.poly.xml|, \verb|town5.sumo.cfg|, \verb|config.xml|), and OMNeT++ scenario definitions (\verb|town5 scenario.ned|, \verb|omnetpp.ini|, \verb|launch.xml|). The \verb|src/| folder contains source code for custom extensions or modules, while the \verb|configure| file helps to set up the simulation environment for compilation. The XML and configuration files define the simulation scenario and simulator configurations. A different scenario suitable for the user application can be implemented simply by changing these files with standard simulation setup files used for these simulators. The order of steps and the files that a user can adjust are illustrated in Figure~\ref{fig:flow-of-sim}.

Firstly, we run the CARLA simulator at the host computer. CARLA simulator follows a server-client architecture. The simulator itself starts as a server. Any number of clients can connect to the server, set up a simulation, run it, and collect necessary data. The clients only need the IP address and port number associated with the CARLA simulator to be able to access the simulator. So, CARLA can also be started on a remote server computer, as long as the client can access the server's IP address and the port is open. The default port for CARLA is 2000. Since we start the CARLA simulator on our local computer, the IP address is that of the local computer, and the port is 2000. If a local computer does not meet the recommended hardware specification in terms of CPU and GPU for CARLA simulator, starting CARLA with lower graphics quality, e.g., \verb|./CarlaUE4.sh -quality-level=Low| may help. The CARLA simulator version 0.9.15 is used in this study.

Then, we start an "Instant-Veins" virtual machine~\cite{instantveins}. The Instant-Veins version provides a pre-built Linux-based environment where the necessary project components are already installed. This project implements co-simulation-related programs and customizations on top of this pre-built environment. The Instant-Veins version \verb|5.2-i1| is used in this study. Since a CARLA simulator is running outside this Instant-Veins virtual machine, we run the virtual machine in "bridged" network adapter mode. A bridged adapter is a network configuration mode used in virtual machines to allow them to be part of the same network as the host computer. This allows the Instant-Veins to be part of the same network as the CARLA server, and thus they can seamlessly connect to CARLA. 

Then, a user clones the GitHub repository in the Instant-Veins environment. From the OMNeT++ IDE,  a user can import the repository as a OMNeT++ project. Since we modify the Veins module of OMNeT++ within the Veins module, it is necessary to clone and import the project from our provided GitHub repository. Then, a user needs to  set the \verb|SUMO_HOME| environment variable to \verb|$HOME/src/sumo| with the command `` \verb|echo `export SUMO_HOME="$HOME/src/sumo"' >> ~/.zshrc|'' and launch the script ``\verb|sumo-launchd.py|'' with the arguments \verb|./sumo-launchd.py -vv -c sumo-gui -n 2|. This will allow connecting to the SUMO simulator. At this point, we can start the co-simulation by right click-menu on the \verb|omnetpp.ini| file. This will launch the \verb|Qtenv| graphical interface, and the simulation can be started by pressing the `Run' icon. This will launch SUMO and connect OMNeT++ as the first client. Later, on a terminal window, a user can run the \verb|run_synchronization.py| script using the arguments specified in the GitHub repository. This example simulation will be running at this stage using Carla Town 05. 

An example command with the necessary parameters for \verb|run_synchronization.py| is as follows.
\begin{verbatim}
  python3 run_synchronization.py --carla-host 10.116.48.5 --sumo-host localhost 
  --sumo-port 49286 --step-length 0.1 --client-order 2 --town-map Town05 
  --tls-manager sumo examples/Town05.sumocfg --debug
\end{verbatim}

It is important to note that the IP address followed by \verb|--carla-host| points to the IP address of the computer where the CARLA simulator is running.

The recommended system configurations and the system configuration used for our development are provided in the Table~\ref{tab:system_config}. 
\begin{table}[htbp]
\centering
\renewcommand{\arraystretch}{1.3}
\caption{System requirements and development configuration}
\begin{tabular}{|l|p{6cm}|p{5cm}|}
\hline
\textbf{Component} & \textbf{Recommended Configuration} & \textbf{System Configuration Used in this Study} \\
\hline
CPU & Intel Core i7 (9th--11th Gen) / Intel Core i9 (9th--11th Gen) / AMD Ryzen 7 / Ryzen 9 & Intel Core i9 (14th Gen) \\
\hline
RAM & 32 GB or more & 64 GB \\
\hline
GPU & NVIDIA RTX 3070 / RTX 3080 / RTX 4090 & NVIDIA RTX 4090 \\
\hline
\end{tabular}
\label{tab:system_config}
\end{table}

\begin{figure}[!htbp]
    \centering
    \begin{subfigure}[b]{0.51\textwidth}
        \centering
        \includegraphics[width=\textwidth]{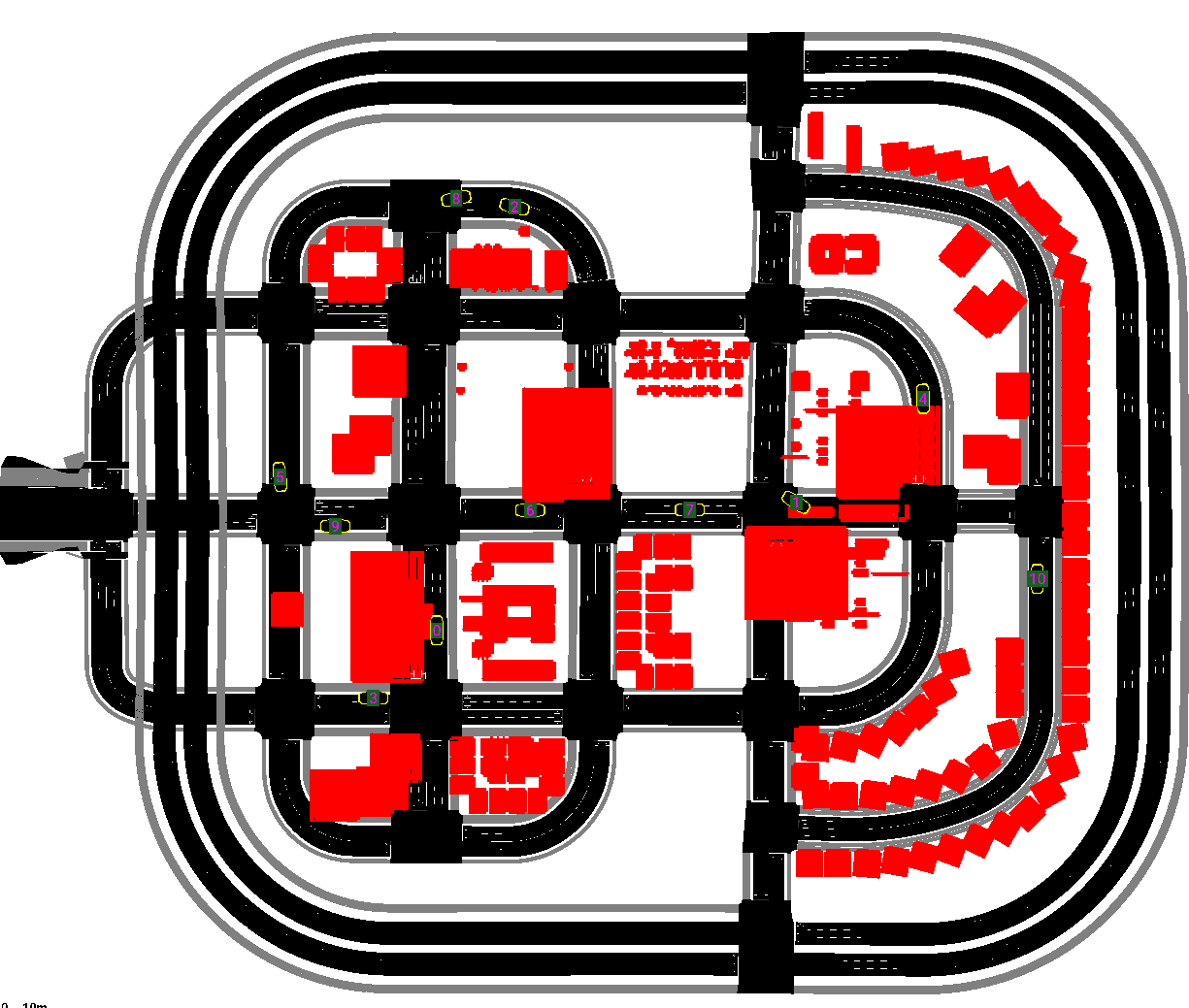}
        \caption{SUMO interface in co-simulation}
        \label{fig:sumo-map}
    \end{subfigure}
    \hfill
    \begin{subfigure}[b]{0.47\textwidth}
        \centering
        \includegraphics[width=\textwidth]{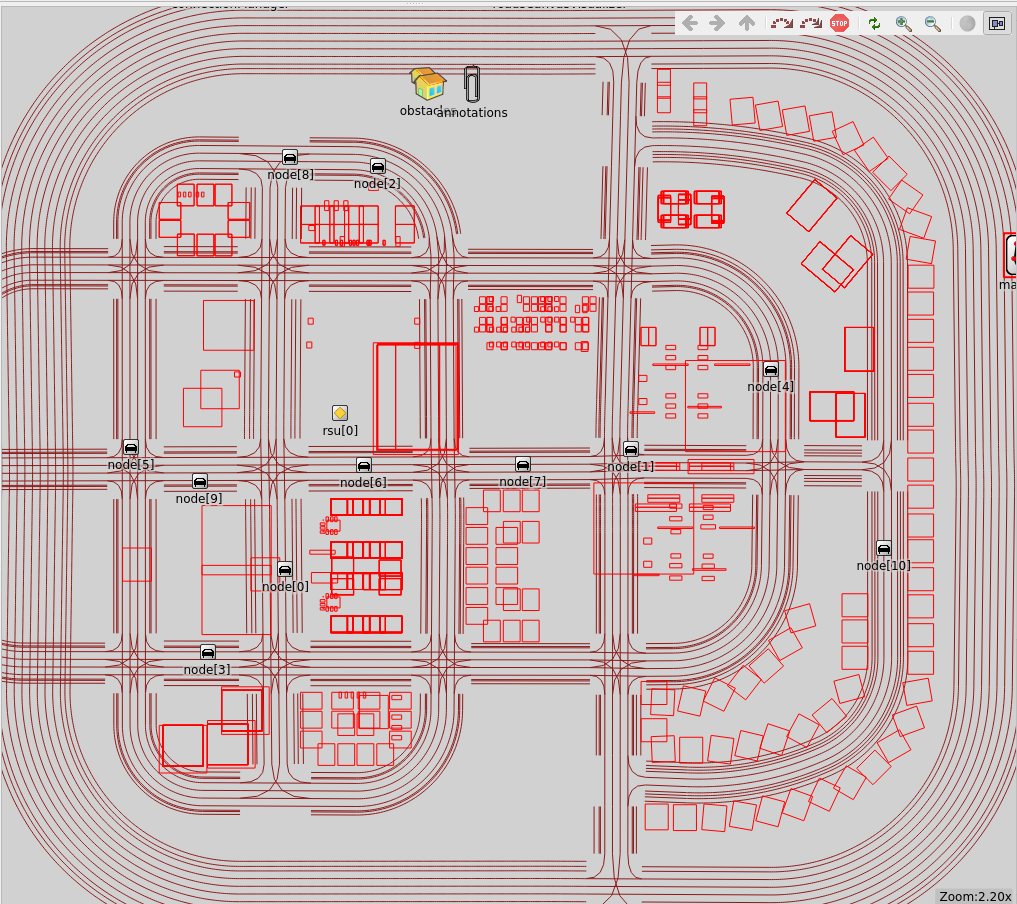}
        \caption{OMNeT++ interface in co-simulation}
        \label{fig:omnet-map}
    \end{subfigure}
    
    \vspace{0.5cm} % Adjust vertical spacing between rows
    
    \begin{subfigure}[b]{1\textwidth} % Wider subfigure for the third image
        \centering
        \includegraphics[width=\textwidth]{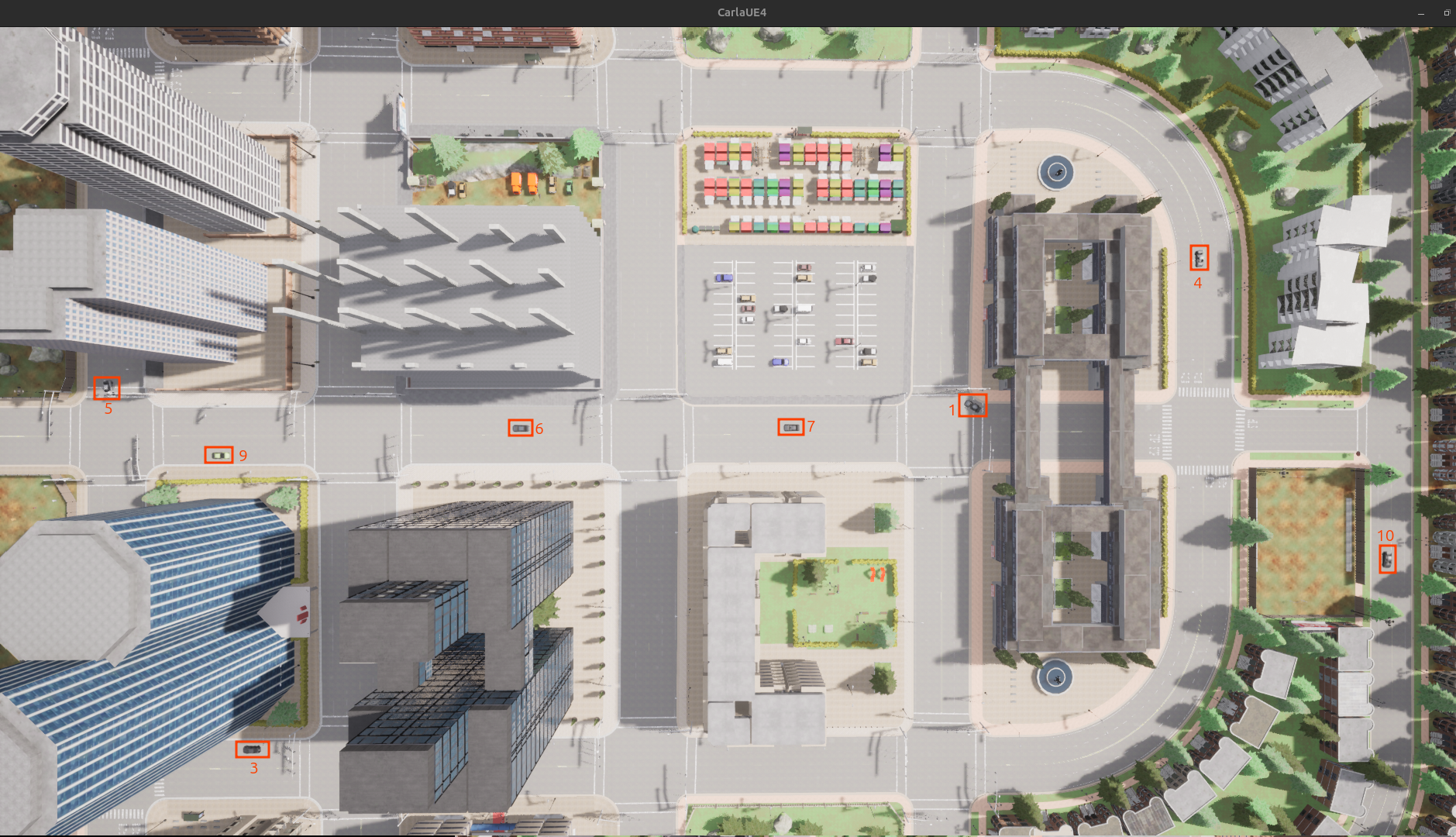}
        \caption{Carla interface in co-simulation}
        \label{fig:carla-map}
    \end{subfigure}
    
    \caption{All three simulators running synchronously.}
    \label{fig:all-sim}
\end{figure}

Figure~\ref{fig:all-sim} illustrates all simulators operating synchronously while executing the same scenario.

\subsection{Collecting data from co-simulation}
\label{data-collection}
CARLA provides a Python client that can be connected to the co-simulation server, provided the IP address and port number. The python client provides the means to interact with the simulation, both collecting and customizing it. The data collection mechanism from CARLA client is no different in the co-simulation environment compared to a standalone CARLA simulation. However, it should be noted that, for maintaining the synchronization between the simulation platform, CARLA is running on a specified delta-second timestamp. The \verb|run_simulation.py| is also connected to CARLA and gives clock ticks at specific times. The user script attached as additional clients should not provide any clock ticks using \verb|world.tick()| function. Only one client should provide the simulation ticks. As such, the data collection client should use a loop with  \verb|world.wait_on_tick()| blocking function. There is no specific limitation on the number of clients that can be connected to the simulation server and collect data or manipulate simulation parameters.

%%% Akid bhai, please review and add to the following paragraph as you see fit. 
OMNeT++ provides a convenient method of storing analytics in scalar (result.sca) and vector files (result.vec). It also provides convenient tools to create visualizations. The INET framework in OMNet++, based on which C-V2X messages are transmitted, provides tooling to capture PCAP (packet capture) recordings, which can be analyzed with popular third-party tools such as Wireshark as well. 

\begin{figure}[ht]
    \centering
    \includegraphics[width=0.4\linewidth]{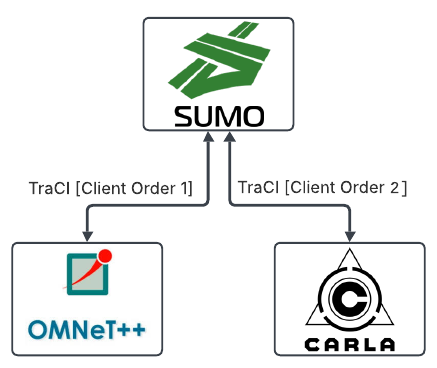}
    \caption{SUMO clients}
    \label{fig:sumo-clients}
\end{figure}

SUMO provides a flexible TraCI interface that allows for the control of all simulation aspects and enables data collection. In the integrated co-simulation environment, the communication with OMNeT++ and CARLA are implemented through the TraCI interface. When starting the co-simulation system, SUMO expects two clients to connect to it. By design, SUMO waits for all clients to connect and set their execution order, as shown in Figure~\ref {fig:sumo-clients}. After all the clients connect, the SUMO simulation can begin. SUMO proceeds to the next simulation step only when all clients (two in this case) request it to do so. In this integrated simulation environment, OMNeT++ is set as client 1, and the program that connects SUMO to CARLA is set as client 2. It is possible to start SUMO with 3 or more clients instead and collect the necessary data. If a user wishes to connect a Python-based TraCI client to collect data from SUMO, the \verb|-n 3| option should be set when running \verb|./sumo-launchd.py| script to accommodate the additional client. The additional client user script should also set its execution order to 3 in the script. An example script that follows this procedure is provided with the GitHub repository under the name - \verb|sumo-additional-client.py|. 

\section{Example Use Cases: Development, evaluation and validation of CAM Applications}
\label{sec:applications}

The OpenCAMS co-simulation platform enables development, evaluation and validation of a diverse and expanding set of applications across CAM systems by synchronizing vehicle dynamics, sensor-level decision-making, and communication networks within a unified simulation loop. Unlike fragmented or domain-specific simulators, OpenCAMS combines SUMO for large-scale traffic modeling, CARLA for high-fidelity vehicle and sensor simulation, and OMNeT++ for application-level wireless communication. This tightly coupled architecture ensures that all simulated behaviors evolve coherently, allowing the framework to accurately capture the complex interactions across cyber and physical layers that are essential to evaluating CAM systems. OpenCAMS supports a broad spectrum of applications, spanning safety, mobility, and cybersecurity, within CAM  systems. Each application fundamentally requires the simultaneous integration of microscopic traffic flow (SUMO), ego-vehicle control and sensor modeling (CARLA), and C-V2X communication (OMNeT++), reinforcing the value of a closed-loop, co-simulation architecture for transportation research.

\textbf{Safety:} OpenCAMS is particularly well-suited for testing safety-critical C-V2X applications that demand immediate and reliable communication. For example, in a Forward Collision Warning use case, an ego vehicle in CARLA receives information about a braking vehicle ahead via OMNeT++, which transmits Basic Safety Messages (BSMs), and must process that information to trigger braking before a collision occurs. Meanwhile, SUMO models the surrounding traffic’s behavior to assess whether secondary collisions or ripple effects are induced by the ego vehicle’s action.  Another example is that protecting Vulnerable Road Users (VRUs), such as pedestrians and cyclists, is a central focus of future transportation safety systems. Using OpenCAMS, researchers can model scenarios in which a pedestrian unexpectedly crosses the road or a cyclist moves into the vehicle’s path. In CARLA, onboard sensors, such as cameras, LiDAR, and radar detect the VRU’s presence. This information is then communicated via C-V2X messages over OMNeT++ to nearby vehicles, enhancing collective situational awareness. SUMO provides the broader context, simulating other vehicles' reactions, pedestrian crossing behavior, and potential congestion buildup. This holistic approach enables safety assessment not only of the ego vehicle’s response but also of how cooperative perception and communication impact the safety of all road users. The system can be evaluated under various visibility conditions, weather scenarios, and communication latencies to ensure robustness. In complex driving environments such as highways or urban arterials, Blind Spot Warning (BSW) and Lane Change Assist (LCA) applications are vital for preventing side-swipe collisions. OpenCAMS facilitates the simulation of these systems by allowing CARLA to model the ego vehicle’s sensor detection of vehicles in adjacent lanes. OMNeT++ transmits relevant Vehicle Status Messages, including position and speed, to assist the ego vehicle in making informed lane-change decisions. Meanwhile, SUMO models the behavior of surrounding traffic, capturing the possible reactions of nearby drivers to the ego vehicle’s lane-change maneuver. Researchers can explore critical safety questions such as whether the warning systems trigger too early or too late, how cooperative vehicles behave versus non-cooperative ones, and how different traffic densities affect the effectiveness of the system.

\textbf{Mobility:} OpenCAMS enables the modeling and evaluation of maneuvers like cooperative merging, adaptive following, and decentralized lane assignments. For instance, when modeling a platoon formation scenario, CARLA simulates each vehicle’s actuation, sensor data processing, and internal decision logic, SUMO controls surrounding traffic flow with vehicles that may or may not be cooperative, and OMNeT++ handles continuous exchange of Cooperative Awareness Messages (CAMs) and other data sharing and coordination packets. This allows the study of how network performance influences controller stability and how non-cooperative background traffic affects maneuver execution. More complex connected infrastructure use cases are also within scope. OpenCAMS supports adaptive signal control, priority-based intersection management, and variable speed limit enforcement. For instance, in a smart corridor scenario, SUMO generates traffic patterns with multiple signalized intersections, CARLA models ego vehicles and perception capabilities, and OMNeT++ simulates Vehicle to Infrastructure (V2I) and Vehicle to Network (V2N) communication for delivering corridor-wide advisories. The interaction of vehicle decisions and network coordination can be analyzed for throughput, fuel economy, and safety compliance. Environmental performance applications, such as Green Light Optimal Speed Advisory (GLOSA), energy-efficient routing, or emissions-based traffic control, require all three domains to be simulated jointly. In GLOSA, SUMO provides signal state transitions and traffic congestion modeling, OMNeT++ delivers SPaT and MAP messages to approaching vehicles, and CARLA models how driver assistance systems or autonomy modules adjust speed and gear settings to minimize fuel use. This simulation structure can be extended to study fleet-level electrification strategies, including cooperative battery management, energy-aware platoon formation, and smart charging station routing. Emergency vehicle coordination and preemption are also supported. A simulated emergency vehicle in CARLA broadcasts its trajectory via OMNeT++, which is received by RSUs modeled in OMNeT++ and processed to update traffic light states in SUMO. The system then evaluates how background traffic reacts, whether nearby connected vehicles make cooperative decisions, and how ego vehicles plan safe responses—all while monitoring end-to-end latency and control performance. Similar approaches can simulate pedestrian safety systems, where virtual pedestrians in SUMO generate crosswalk events, which are sensed by infrastructure (e.g., cameras or radar in CARLA) and used to trigger V2P warnings via OMNeT++.

\textbf{Cybersecurity:} OpenCAMS also provides a rigorous platform for cybersecurity testing.  In a Sybil attack, an adversary injects multiple fake identities into the communication network to mislead other vehicles. In OpenCAMS, OMNeT++ simulates the transmission of falsified BSMs or cooperative perception messages from non-existent vehicles.
The ego vehicle in CARLA, receiving these messages, may incorrectly interpret the traffic density around it as high, prompting it to brake unnecessarily or change lanes to avoid a phantom threat. This false reaction can result in rear-end collisions or lane drift conflicts, particularly in dense traffic scenarios modeled in SUMO. Researchers can vary the number, position, and transmission timing of the fake vehicles to evaluate system vulnerabilities and test detection algorithms. In a replay attack, valid messages are captured and maliciously retransmitted at a later time. In OpenCAMS, OMNeT++ can simulate the replay of outdated or context-inappropriate BSMs, such as a vehicle claiming to be stopped at an intersection when it is actually no longer there. The ego vehicle in CARLA may respond to these false messages by slowing down, rerouting, or taking evasive actions that are no longer warranted, introducing unnecessary delays and safety risks. SUMO captures the systemic consequences, including the formation of congestion clusters, reduced throughput, or erratic vehicle behaviors. This allows researchers to test message freshness verification mechanisms, timestamp validation, and authenticated message protocols in real-time. In this scenario, an attacker injects malicious control commands into the infrastructure, specifically targeting traffic signals managed in SUMO. By manipulating the signal phases—e.g., keeping all directions green or causing frequent toggling—the attacker can create unsafe intersection conditions. The ego vehicle in CARLA, relying on SPaT data transmitted through OMNeT++, may proceed through an intersection at the wrong time, resulting in potential cross-traffic collisions. OpenCAMS allows researchers to simulate these attacks with controlled timing and evaluate authentication mechanisms, signal integrity checks, and vehicle-based cross-verification techniques using onboard sensors or crowd-sourced data.

\textbf{Digital Twin (DT) for safety, mobility and cybersecurity} The framework is also powerful for developing digital twin systems, which require a virtual replica of both traffic behavior and communication flow to guide real-time decisions, for safety, mobility and cybersecurity research. For instance, an urban intersection can be modeled with background traffic generated by SUMO, high-resolution ego vehicle behavior captured by CARLA, and roadside units (RSUs) or cloud services communicating via OMNeT++. The twin can then predict downstream traffic conditions, dynamically alter signal timing, or detect traffic anomalies due to abnormal behavior or infrastructure faults. A Digital Twin of a transportation system (e.g., road network, connected vehicles, traffic control infrastructure) can simulate cyberattacks in a controlled virtual environment. Simulating a GPS spoofing attack on an autonomous vehicle to study how it affects traffic flow, safety, and vehicle behavior. Helps identify vulnerabilities and improve incident response plans without real-world risks. DTs allow the safe testing of cybersecurity solutions, such as intrusion detection systems (IDS) for in-vehicle networks or anomaly detection algorithms for C-V2X communication. For example, virtually inject malware into a DT of an in-vehicle network (e.g., CAN bus) to evaluate how well the IDS can detect and respond to the attack. Fine-tunes security solutions before deployment in the field. By continuously mirroring the physical transportation system, the DT can detect deviations between expected (simulated) and actual behavior. For example, an unexpected sudden brake command or irregular communication pattern in a vehicle’s DTcould flag a cyber intrusion attempt. This enables real-time cyber threat detection with predictive capabilities to prevent attacks before damage occurs.

Overall, OpenCAMS’s architecture encourages extensions to emerging research frontiers, such as human-autonomy interaction (by simulating mixed traffic in SUMO and hybrid control in CARLA), teleoperation under network constraints (via uplink/downlink simulation in OMNeT++), and integration with city-scale digital infrastructure through cloud simulation proxies. Python clients can dynamically alter infrastructure behavior or insert synthetic data streams for perception fusion experiments. It is important to note that OpenCAMS facilitates scalable scenario-based benchmarking. Researchers can configure high-traffic-density freeway segments, urban grids with complex signal timing, or adverse weather conditions in CARLA to benchmark perception performance under fog or rain. All such tests benefit from synchronized network modeling and background traffic conditions to validate performance trade-offs. Benchmarks can target algorithms for trajectory prediction, misbehavior detection, consensus control, or cooperative lane selection, enabling reproducible comparisons under tightly controlled settings.

%For example, communication threats, such as Sybil attacks, replay attacks, or GNSS spoofing can be modeled in OMNeT++, with CARLA assessing how compromised data alters vehicle planning, and SUMO tracking the consequences on network-level safety and mobility. For example, a Sybil attack injecting multiple fake vehicles into the communication network can lead an ego vehicle to perform unnecessary braking in CARLA, which may cause rear-end collisions and bottlenecks in SUMO. OpenCAMS allows injection of adversarial packets with timing control and supports measuring latency, packet loss, and throughput under varying attack strategies. GNSS simulators can introduce position spoofing or jamming effects on the ego vehicle, further enabling cyber-physical threat modeling. 

%\section{Case Study}
% cv2x based collision warning system
% task 1: make a diagram data collect (carla lidar), prediction (LSTM), warning (with PCQ),

% task 2: show prediction accuracy, show near miss (BEV), cv2x related matrix (akid bhai)

% Figure: wall clock time, simulation time in each simulator. 

% define scenario
% define output
\section{Conclusions}
\label{sec:conclusions}

We introduced OpenCAMS, a synchronized co-simulation platform integrating SUMO, CARLA, and OMNeT++, to model CAM systems for transportation safety, mobility and cybersecurity research. By enabling discrete-time step synchronization across traffic dynamics, vehicle-level perception and control, and C-V2X communication, OpenCAMS provides a comprehensive platform for developing, evaluating, and validating a wide range of transportation research within a transportation cyber-physical system and/or transportation digital twin. Its design ensures closed-loop fidelity across domains, allowing researchers to explore cross-layer dependencies with precise control over roadway network dynamics, sensor capabilities, and mobility behavior. The use of widely supported open-source tools, coupled with an expandable architecture, makes OpenCAMS a flexible and reproducible testbed for transportation research.

Future research will focus on extending OpenCAMS's capabilities to support emerging technologies and research domains. For example, the integration of 5G NR sidelink protocols, ROS2-based autonomy stacks, and real-time GNSS spoofing models will enable the framework to address emerging C-V2X scenarios and cyberattacks on localization. Another example could be the incorporation of infrastructure-based intelligence via multi-access edge computing (MEC), as well as the streaming of real-world sensor or map data for digital twin validation. This will expand OpenCAMS's utility for infrastructure operators and city planners. Moreover, advancing the platform’s cyber-physical resilience capabilities by embedding formal security models, trust management systems, and certificate validation mechanisms will enable researchers to simulate and mitigate threats at scale. As CAM system continues to evolve, OpenCAMS stands as a critical enabling tool for exploring safe, intelligent, efficient, and secure mobility solutions under tightly integrated, realistic simulation conditions.

\bibliographystyle{unsrt}  
\bibliography{cosimulation-article}

\end{document}